\documentclass[aps,prb, reprint, superscriptaddress]{revtex4-2}
\usepackage{graphicx}% Include figure files
\usepackage{dcolumn}% Align table columns on decimal point
\usepackage{bm}
\usepackage{nicefrac}
\usepackage{tabularx}

\usepackage{floatrow}
\usepackage{siunitx}

\usepackage{amsmath}

\usepackage{amsmath,amssymb,bm}
\usepackage{graphicx}
\usepackage{xspace}
\usepackage{latexsym}
\usepackage{color}
\usepackage{hyperref}
\usepackage{placeins}

\usepackage{soul}

\newcommand{\SrZn}{SrZnVO(PO$_4$)$_2$\xspace}

\newcommand{\BaCd}{BaCdVO(PO$_4$)$_2$\xspace}

\newcommand{\PbV}{Pb$_2$VO(PO$_4$)$_2$\xspace}

\newcommand{\be}{\begin{equation} }
\newcommand{\ee}{\end{equation} }
\newcommand{\bea}{\begin{eqnarray} }
\newcommand{\eea}{\end{eqnarray} }

\newcommand{\mb}[1]{\mathbf{#1}}

\newcommand{\K}{\kelvin}
\newcommand{\T}{\tesla}

\newcommand{\meV}{\milli\electronvolt}
\newcommand{\VV}{V$^{4+}$\xspace}

\begin{document}

%\preprint{}

\title{Phase diagram and spin waves in the frustrated ferro-antiferromagnet \SrZn }


\author{F. Landolt}
\email[]{landoltf@phys.ethz.ch}
\author{Z. Yan}
\author{S. Gvasaliya}
\affiliation{Laboratory for Solid State Physics, ETH Zürich, 8093 Zürich, Switzerland}

\author{K. Beauvois}
\author{E. Ressouche}
\affiliation{Université Grenoble Alpes, CEA, IRIG, MEM, MDN, 38000 Grenoble, France}

\author{J. Xu}
\homepage{current address: Heinz Maier-Leibnitz Zentrum (MLZ), Technische Universit\"{a}t M\"{u}nchen, 85748 Garching, Germany}
\affiliation{Helmholtz-Zentrum Berlin für Materialien und Energie, Berlin, Germany}

\author{A. Zheludev}
\homepage{http://www.neutron.ethz.ch/}
%\thanks{}
\affiliation{Laboratory for Solid State Physics, ETH Zürich, 8093 Zürich, Switzerland}


\date{\today}

\begin{abstract}
Single crystals of the frustrated $S=1/2$ ferro-antiferromagnetic proximate square lattice material \SrZn are studied in magnetometric, calorimetric, neutron diffraction and inelastic neutron scattering experiments. The measured spin wave spectrum reveals a substantial degree of magnetic frustration and a large quantum renormalization of the exchange constants. The $H-T$ magnetic phase diagram is established. It features a novel pre-saturation phase, which appears for only one particular field orientation. The results are discussed noting the similarities and differences with the previously studied and similarly structured \PbV compound.
\end{abstract}

%\keywords{}

\maketitle

% body of paper here - Use proper section commands pyroglove
\section{Introduction} \label{sec:Intro}

Layered vanadyl phosphates with the formula ABVO(PO$_4$)$_2$ (A,B = Sr, Zn, Pb, Ba, Cd) \cite{Nath2008,TsirlinSchmidt2009,TsirlinRosner2009} have attracted considerable attention, as they realize the frustrated ferro-antiferromagnetic $S=1/2$ Heisenberg model in an (approximate)  $J_1-J_2$  square-lattice geometry. The interest stems from theoretical predictions of exotic quantum magnetic states that may appear in such systems for either strong frustration or close to saturation in external magnetic fields. In the latter case, presaturation phases are expected to occur above the field of single-magnon Bose-Einstein condensation and below full polarization at $H_\mathrm{sat}$ \cite{Ueda2013}.  Specifically, the ideal square-lattice version of the model features a quantum spin nematic phase \cite{Shannon2006,Shindou2009,Smerald2015} that is particularly robust in applied fields \cite{Ueda2013,Ueda2015}. 

To date, unconventional pre-saturation phases have been identified in two species, namely \BaCd \cite{Bhartiya2019} and \PbV \cite{Landolt2020}. The former compound is highly frustrated \cite{Povarov2019,Bhartiya2021} with a much reduced ordered moment in zero field \cite{Skoulatos2019}, and a very broad pre-saturation phase. This phase appears in all field orientations and clearly lies above the single-magnon instability field \cite{Bhartiya2019,Bhartiya2021}. All the information available to date is consistent with it being a quantum spin-nematic. In contrast, quantum fluctuations in \PbV are much weaker, the zero-field ordered moment is quite large \cite{Bettler2019}, and the presaturation phase is very narrow \cite{Landolt2020}. Whether it lies above single-magnon condensation field is yet unclear. Unlike in \BaCd, it appears only in magnetic fields applied perpendicular to the easy axis of weak magnetic anisotropy in this material \cite{Landolt2020}.  With certainty, it is not a pure spin-nematic, as NMR experiments unambiguously show the presence of (perhaps rather complex) dipolar order and/or phase separation \cite{Landolt2020}.

Is any of this behavior generic, or does it arise from the specific rather complicated and substantially different interaction geometries in the two systems? Seeking more insight into this question, we study \SrZn, a third member of the family. It is believed to be intermediate between the above-mentioned species in terms of magnitude of frustration and the magnitude of quantum fluctuations \cite{Tsirlin2009,Bossoni2011,Skoulatos2009}. We use inelastic neutron scattering to map out spin wave spectra in zero field and determine the relevant exchange constants. In stark contrast with \PbV, in \SrZn we find a strong quantum renormalization of the spin wave energies. Neutron diffraction also reveals an ordered moment that is considerably more suppressed by quantum fluctuations  than in the Pb-based compound. Nevertheless, magnetic and thermodynamic measurements that we employ to map out the $H-T$ phase diagram uncover a narrow pre-saturation phase almost identical to the one seen in \PbV and clearly distinct from that in \BaCd.

\begin{figure*}
	\includegraphics[width=\columnwidth]{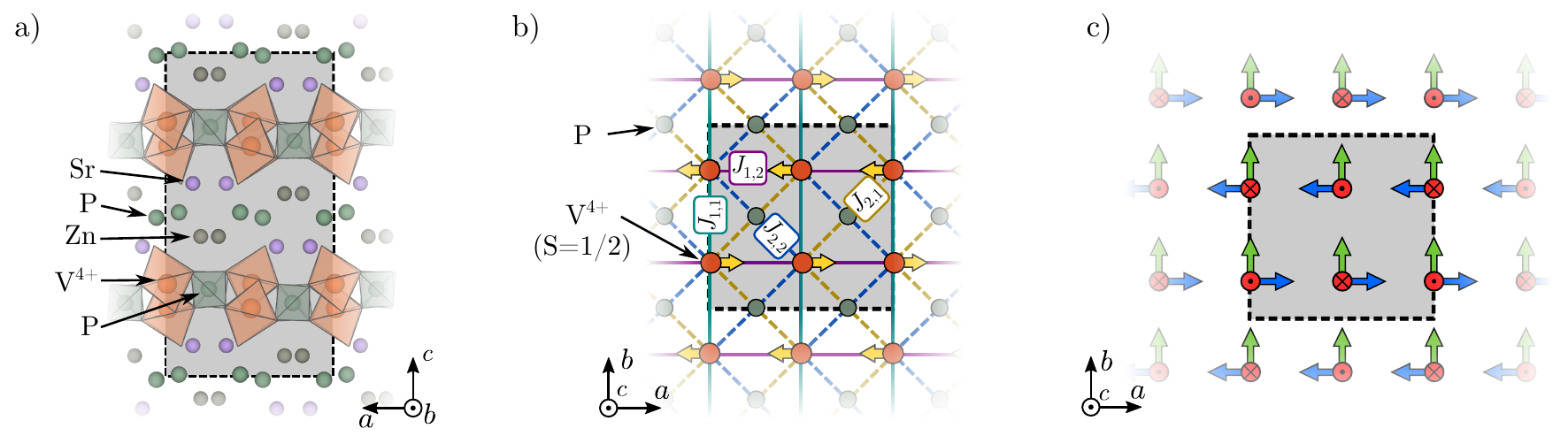}
	\caption{\label{fig:struct} Schematic view of the \SrZn crystal structure. (a) Projection of the entire structure onto the $(a,c)$ plane. Shown are the VO$_5$ pyramids, the PO$_4$ tetrahedra, as well as the Sr and Zn sites. Oxygen atoms are omitted for clarity. Shaded rectangles indicate the crystallographic unit cell. (b) A single vanado-phosphate layer in the $(a,b)$ crystallographic plane. Lines represent exchange bonds between the $S=1/2$ \VV ions, as described in the text. In the ordered phase, the magnetic structure is a columnar antiferromagnet (CAF) with spins pointing along  $\mb{a}$ and alternating along  $\mb{b}$, as indicated by yellow arrows. The layers are ferromagnetically correlated along  $\mb{c}$. (c) Illustration of the \VV local environment. It shows how the crystal symmetry operations transform a local quiver between the vanadium sites.}
\end{figure*}

\begin{figure*}
	\includegraphics[width=\columnwidth]{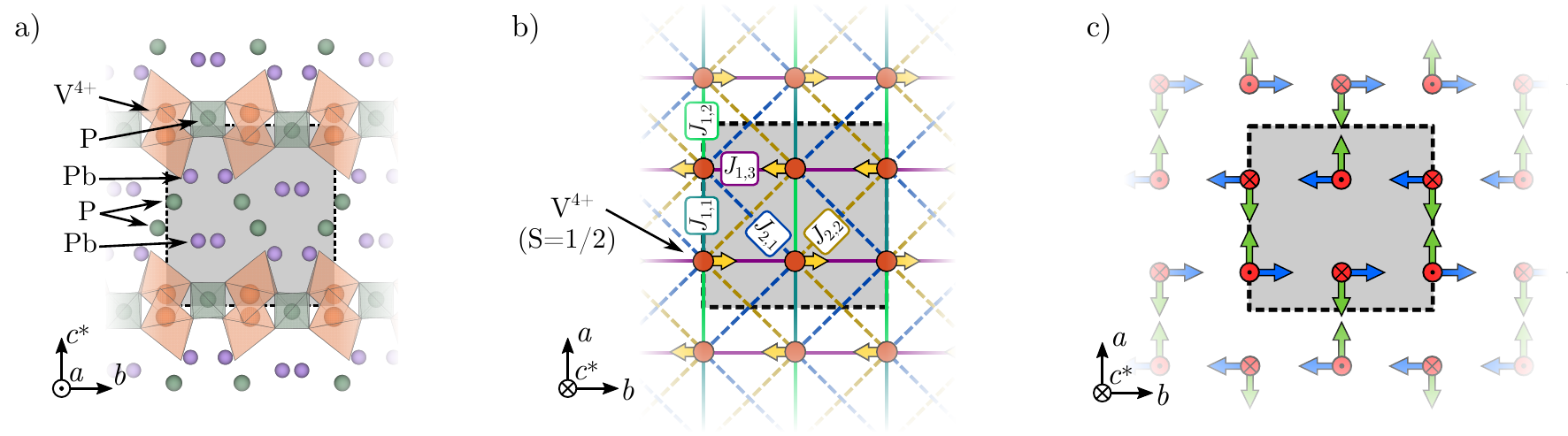}
	\caption{\label{fig:struct_Pb} Same as Fig.~\ref{fig:struct}, but for \PbV.}
\end{figure*}

The crystal structure of \SrZn is shown in Fig.~\ref{fig:struct} ~a). The unit cell is orthorhombic (Pbca) with lattice parameters $a = \SI{9.066(1)}{\angstrom}$, $b=\SI{9.012(1)}{\angstrom}$ and $c = \SI{17.513(1)}{\angstrom}$ \cite{SrZn_Struct} at room temperature. The magnetism originates from $S=1/2$ V$^{4+}$ ions that form layers in the $(a,b)$ plane in {\em approximately} a square-lattice geometry. There are 8 crystallographically equivalent V$^{4+}$ sites, 4 in each layer. Magnetic interactions within the layers are through  V-O-P-O-V superexchange pathways and are illustrated in Fig. \ref{fig:struct}~b). Symmetry allows for two nearest-neighbor (nn) coupling constants $J_{1,1}$ and $J_{1,2}$ along the crystallographic $\mb{a}$ and $\mb{b}$ axes, respectively. There are also two distinct next-nearest neighbor (nnn) interactions  $J_{2,1}$ and $J_{2,2}$, along the diagonals of the proximate square lattice. 

Of the three vanadyl phosphates mentioned here, \SrZn appears to be the simplest one in terms of structure and the number of distinct magnetic interactions. Magnetic coupling between layers is assumed to be negligible, based on the previous direct measurements in \PbV \cite{Bettler2019}.

\section{Experimental} \label{sec:Exp}

\begin{figure*}
	\begin{floatrow}	
		\ffigbox[\FBwidth]{\includegraphics[width=8.67cm]{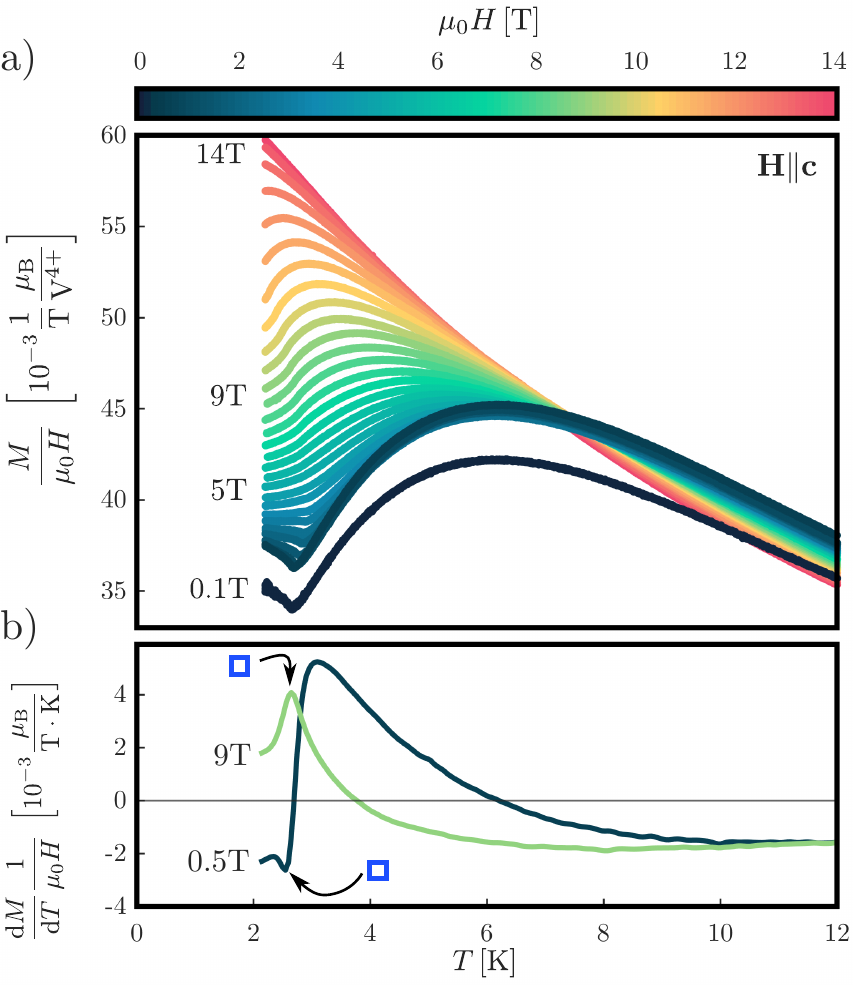}}{
			\caption{(a) Magnetization data measured in temperature scans at constant fields along $\mathbf{c}$. (b) Representative temperature derivatives of two magnetization scans show in (a). Extrema in the derivatives are associated with the onset of long-range magnetic order. The symbols correspond to the phase boundary points in Fig.~\ref{fig:PhaseDiagram} (a).  \label{fig:VSM}} }
		
		\ffigbox[\FBwidth]{\includegraphics[width=8.67cm]{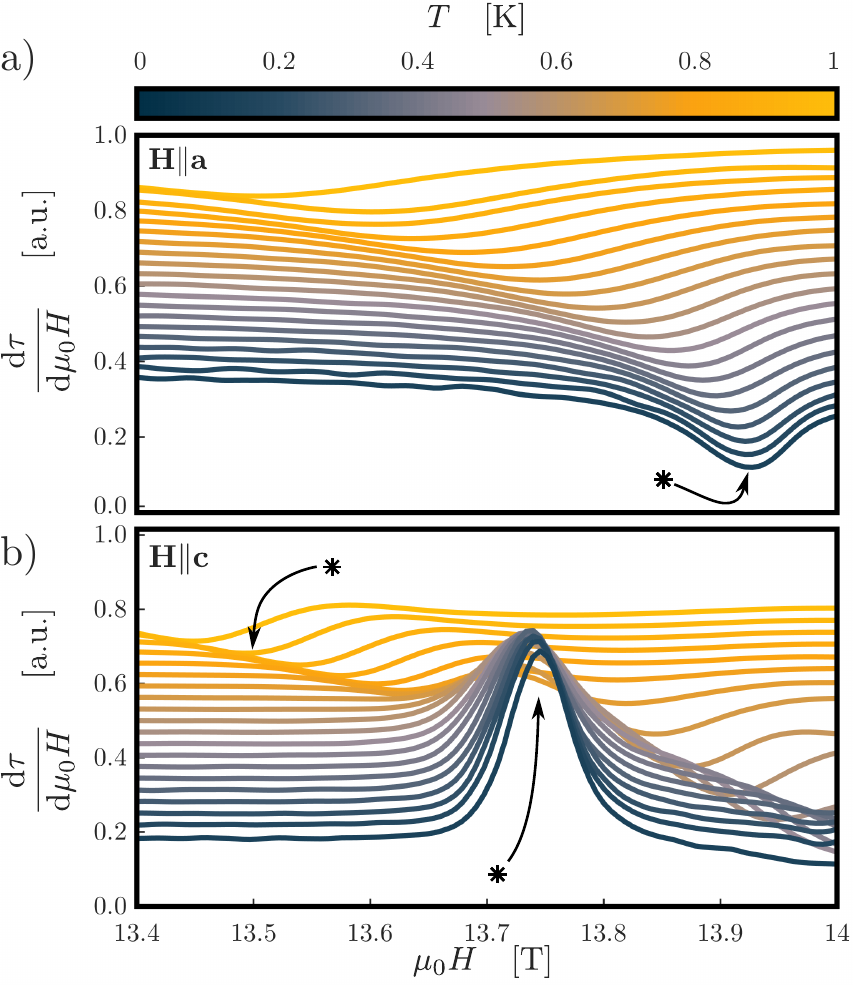}}{
			\caption{ Field derivatives of magnetic torque scans for two field geometries, $\mb{H}\|\mb{a}$  (a) and $\mb{H}\|\mb{c}$  (b). Individual curves are offset vertically to be equally spaced at \SI{13}{\T} for visibility. Peaks in the torque derivative are associated with phase boundaries (asterix).  \label{fig:torque}	} }
	\end{floatrow}
\end{figure*}

\begin{figure}
	\includegraphics[width=8.67cm]{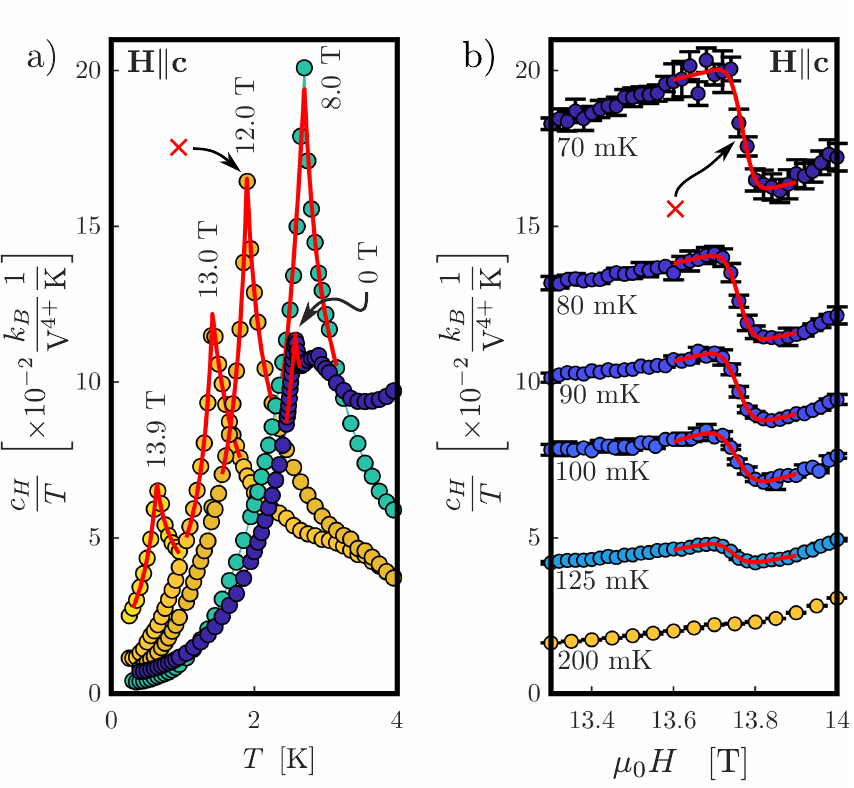}
	\caption{Temperature scans (a) and field scans (b) of specific heat measured in \SrZn in magnetic fields applied along the $\mb{c}$ axis. Solid lines are empirical fits to the data to determine the corresponding transition points, as described in the text.\label{fig:HeatCap}}
\end{figure}

Single crystal samples of \SrZn were grown by the self-flux Bridgman technique in a quartz crucible sealed under a vacuum of $10^{-4}$~torr. A small piece of tantalum metal was used as oxygen scavenger.

Two geometries, with the external magnetic fields applied parallel to the crystallographic $\mb{a}$ (magnetic easy axis) and $\mb{c}$ axes were explored using bulk measurements.
Magnetization data were collected using a Quantum Design (QD) vibrating sample magnetometer (VSM) PPMS insert on a $6.4$~mg sample. Measurements were done by scanning temperature at fixed fields. Additional field scans for fields along $\mb{a}$ were measured on a QD MPMS reciprocating sample option (RSO) using a \SI{6.0}{\milli\gram} sample. A custom Faraday force magnetometer \cite{Blosser2020} PPMS insert was used to measure magnetic torque down to dilution temperatures on a $0.2$~mg sample. The same sample was used in heat capacity measurements using a QD PPMS with a heat-pulse calorimeter option and a dilution refrigerator insert.

Neutron diffraction experiments were carried out on the CEA-CRG D23 lifting-counter diffractometer at ILL using neutrons with a wavelength of $\lambda = 2.35$~\AA. An 80~mg single crystal sample was mounted in a dilution cryostat with the $\mb{c}$ axis vertical. Sample mosaic was 0.5°. Higher order beam contamination was suppressed using a pyrolitic graphite (PG) filter. The experiment was carried out in zero external magnetic field. The low-temperature crystal structure was verified at $T=\SI{10}{\kelvin}$ by measuring the intensities of 126 nuclear Bragg reflections. Magnetic ordering was detected by the appearance of the otherwise-forbidden (0,1,0) Bragg reflection upon cooling below $T_\mathrm{N}=\SI{2.6}{\kelvin}$. The emerging magnetic structure corresponds to a $(0,0,0)$ propagation vector, so that most magnetic Bragg peaks overlap with nuclear ones. Overall, intensities of 65 potentially magnetic Bragg peaks were measured first in the ordered phase at 50~mK, then in the paramagnetic state at  10~K. The data were collected in rocking scans. Each scan consists of 35 points and a counting time of about 3~s/point. To improve statistic, the scattering intensity around each peak position was counted for about 25 minutes. 
The difference between data sets at the two temperatures was attributed to magnetic scattering. The underlying magnetic structure was refined using FullProf \cite{RodriguezCarvajal1993}.

Inelastic neutron scattering was carried out on the cold-neutron triple axis spectrometer FLEXX at Helmholtz-Zentrum Berlin (HZB) \cite{Le2013}. Two crystals of total mass 867~mg were co-aligned to a cumulative mosaic of 1.8°. Sample environment was a standard ``orange'' cryostat with an dilution insert but no magnet was employed. Mounting the sample with $\mb{c}$ vertical enabled access to momentum transfers in the  $(h,k,0)$ reciprocal-space plane.   All measurements were performed with a fixed final energy of 5~meV, a double focusing PG monochromator and horizontal focusing PG analyzer. Higher-order beam contamination was reduced by using a Be filter and a velocity selector. The measured energy resolution at the elastic position was 0.3~meV full width at half maximum. The data was collected in energy scans, typically counting 5~min. per point.

\begin{figure*}
	\includegraphics[width=18cm]{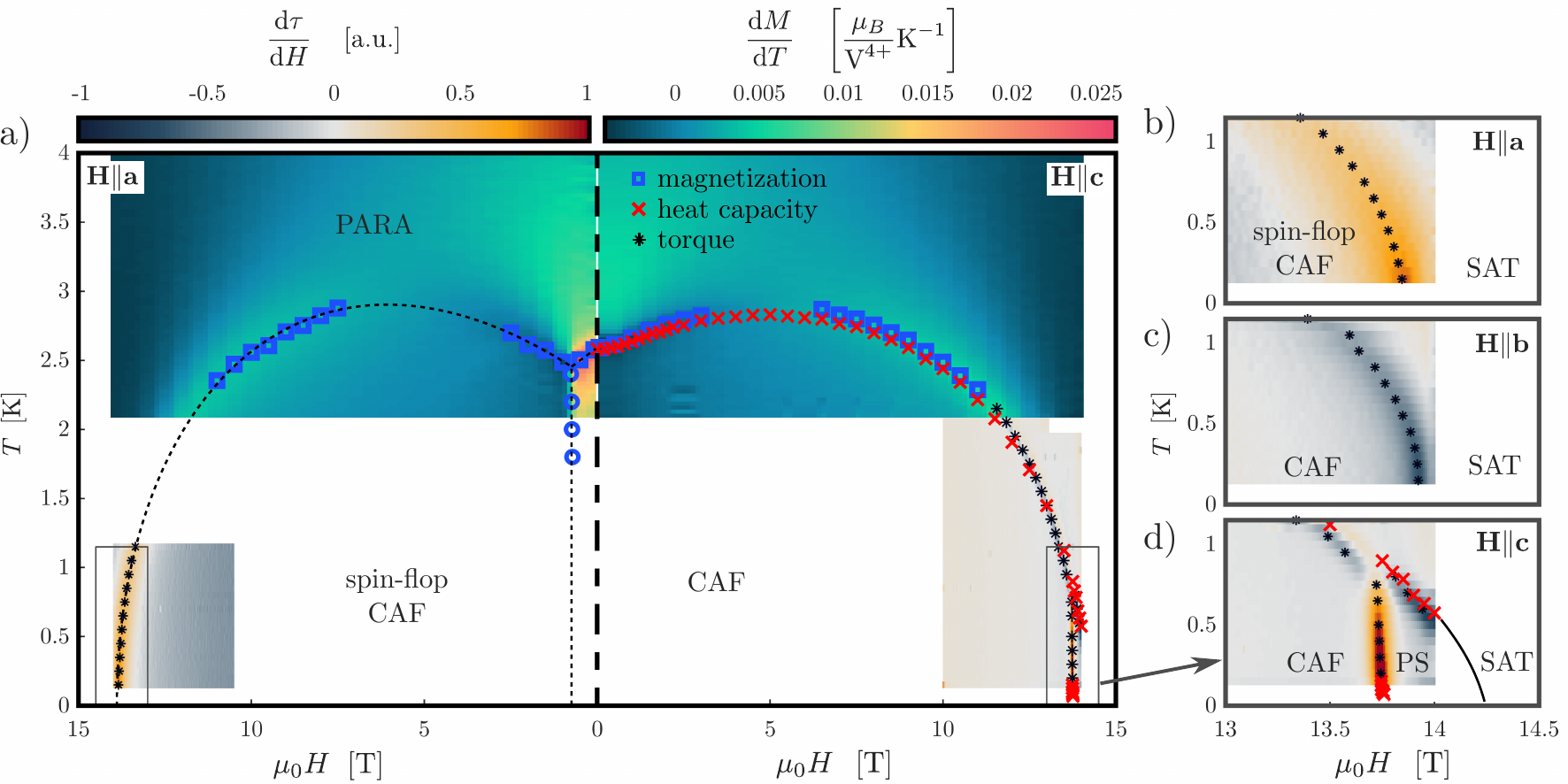}
	\caption{\label{fig:PhaseDiagram} Magnetic phase diagram of \SrZn for different field orientations. The false-color panels correspond to either temperature derivative of magnetization or the field derivative of magnetic torque. The black dashed line indicates  boundaries separating the paramagnetic (PARA), columnar antiferromagnetic (CAF) and pre-saturation (PS) phases. Black dots, squares and crosses are transition points determined from torque, magnetization and specific heat data, respectively, as described in the text. }
\end{figure*}

\section{Results} \label{sec::ExpRes}

\subsection{Magnetic phase diagram}
As was previously done for \PbV \cite{Landolt2020}, measurements of magnetization vs. temperature at a constant value of applied field  were used to map out the magnetic phase diagram in \SrZn down to \SI{2.1}{\kelvin}. The measured $M(T)$ curves for $\mb{H}\|\mb{c}$ are shown in Fig.~\ref{fig:VSM} along with two representative derivative plots. Arrows indicated features associated with the onset of magnetic long range order  \cite{Landolt2020}. False-color plots of the derivatives are shown in the top part of Fig.~\ref{fig:PhaseDiagram}~a). Open blue squares indicate the phase transitions identified. Field scans of magnetization for $\mb{H}\|\mb{a}$ show a clear sign of a spin-flop  at $\mu_0 H_\mathrm{SF}\sim \SI{0.73}{\tesla}$, indicated by open blue circles in  Fig.~{\ref{fig:PhaseDiagram}~a). 

The low-temperature-high field  sections of the phase diagram were mapped out using magnetic torque experiments.  The torque signal $\tau$ was measured in field scans at fixed temperatures. Field-derivatives of $\tau$ are shown in Fig.~\ref{fig:torque} and in false color plots in Fig.~\ref{fig:PhaseDiagram}~b)-d). For $\mb{H}\|\mb{a}$ there appears to be a single feature (asterisk) in the data, indicative of a phase transition. We associate this feature with a transition to the fully saturated paramagnetic phase. An extrapolation of its temperature dependence to $T=0$ yields $\mu_0 H_{\mathrm{sat},a}=13.92$~T.

\begin{figure*}
	\includegraphics[width=18cm]{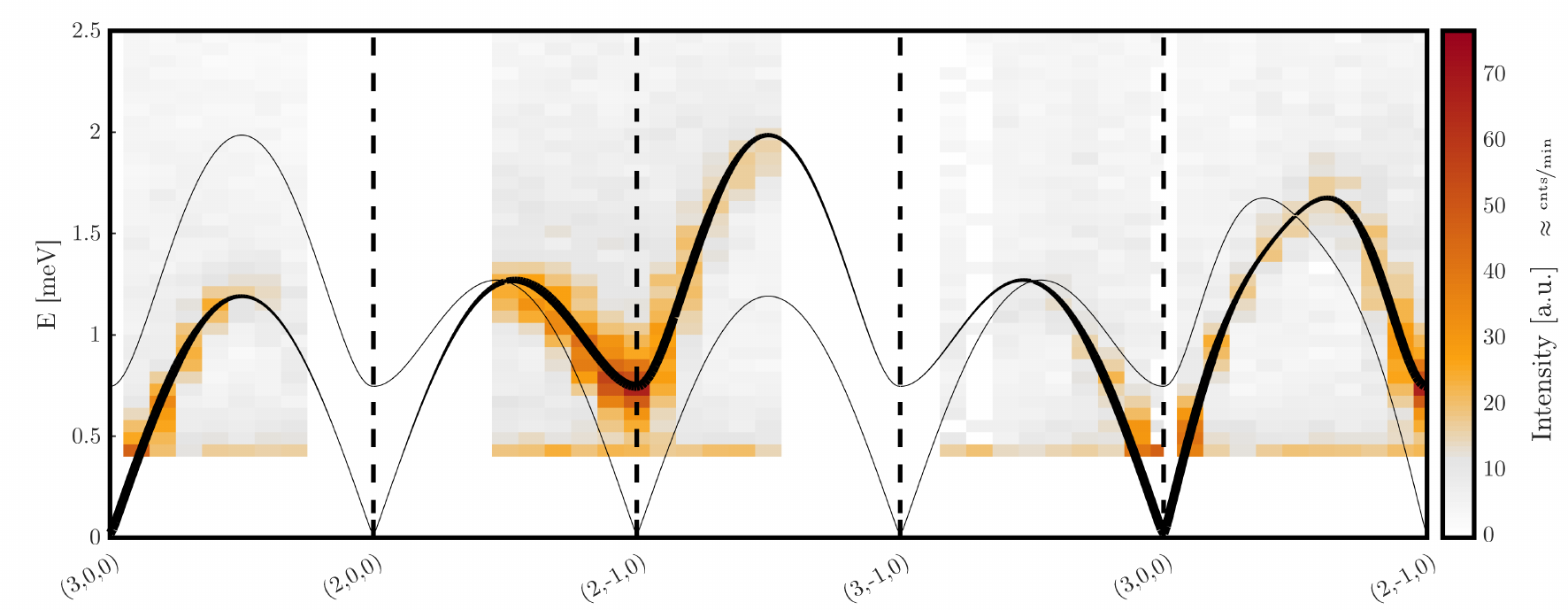}
	\caption{\label{fig:Neutron} False color plot of inelastic neutron scattering intensities measured in \SrZn at $T=0.1$~K. Solid lines are the spin wave dispersion calculated using the exchange couplings of Tab.~\ref{tab:J}, as obtained in a global fit to the data. Their widths correspond to the calculated intensities.}
\end{figure*}

\begin{figure}
	\includegraphics[width=8.67cm]{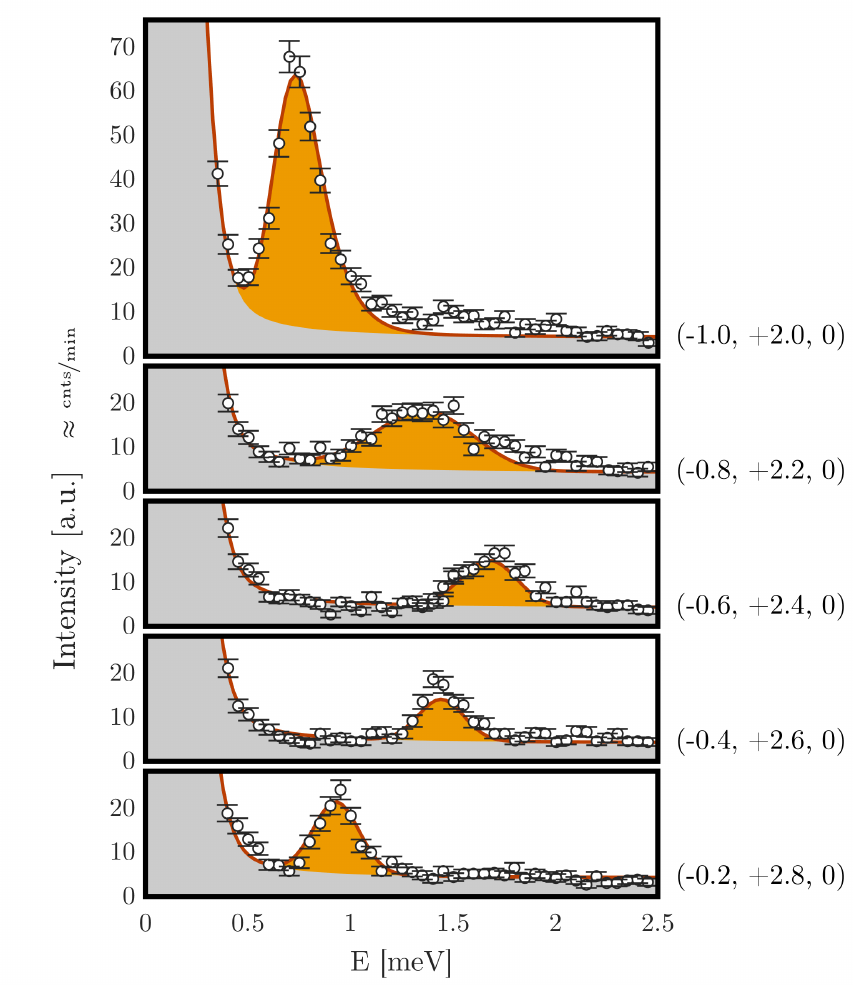}
	\caption{\label{fig:Neutron_Scan} Symbols: representative constant-$Q$ inelastic neutron scattering spectra measured in \SrZn at $T=0.1$~K. Shaded areas correspond to the calculated background (grey) and magnetic scattering (orange), obtained by a global fit of a spin wave model to all 49 measured scans.}
\end{figure}

The torque signal for $\mb{H}\|\mb{c}$ shows a qualitative similar behavior above about 0.7~K. At lower temperature however, an additional feature appears at $\mu_0 H_c=13.75$~T. The position of this new  feature is almost temperature-independent and becomes only sharper at lower temperatures. We associate it with the appearance of an additional pre-saturation phase. The second feature at full saturation retains its high temperature character and moves outside the accessible field range at low temperature. The temperature dependence of $\mu_0 H_{\mathrm{sat},c}(T)$ appears to be compatible with the previous reported saturation field of $\mu_0 H_{\mathrm{sat, powder}}=\SI{14.2}{\T}$ determined in powder experiments \cite{Tsirlin2009}. Overall, the observed behavior of magnetization and torque in magnetic fields applied parallel or perpendicular to the anisotropy axis is remarkably similar to that previously seen in \PbV \cite{Landolt2020}. 

The accessibly low fields make it possible to confirm the phase diagram with specific heat measurements. Typical temperature scans of specific heat at constant external fields applied along $\mb{c}$ are shown in Fig.~\ref{fig:HeatCap} a) (symbols). The boundary of the paramagnetic phase  is marked by a distinct $\lambda$-anomaly. Empirical power-law fits (solid lines in Fig.~\ref{fig:HeatCap}~a) ) were used to determine the transition field, similar to the procedure described in \cite{Huevonen2012}. Also the transition to the pre-saturation phase becomes visible at the lowest temperatures. As shown in Fig.~\ref{fig:HeatCap}~b), it manifests itself as a step in the measured field scans. Its magnitude decreases with increasing temperature. Above 200~mK the feature is no longer visible. The transition field was associated with the inflection point in the data and obtained by empirically function fit (solid lines in Fig.~\ref{fig:HeatCap}~b) ) \footnote{	$c_H^{jump}(H) = A\cdot \text{erf}\left(a\cdot(H-H_c)\right) + b\cdot H + c$ with free parameters $A,a,H_c,b$ and $c$.}. The resulting phase boundary data for both transitions are shown by red crosses in Fig.~\ref{fig:PhaseDiagram}.

\subsection{Diffraction}
An analysis of 126 nuclear Bragg intensities measured at $T=10$~K revealed that the crystal structure is essentially the same as at room temperature, space group Pbca and lattice parameters $a=9.07(1)$~\AA, $b=9.00(1)$~\AA ~and $c=17.44(2)$~\AA.

For the magnetic structure refinement a $q=0$ ordering vector was assumed with spins pointing along $\mb{a}$. In the representation analysis this ordering vector and the space group Pbca, leads to eight irreducible representations \cite{Wills2000}. The best agreement with the measured Bragg intensity was obtained by the one corresponding to the CAF structure,  depicted in Fig.~\ref{fig:struct}~b). The weighted R-factor is 19.9\%. Each layer has the same magnetic structure and are therefore ferromagnetically correlated. The ordered moment is $0.361(6)~\mu_\mathrm{B}$ per V$^{4+}$ site and could be determined by using the scale factor from the structural refinement at \SI{10}{\K}. This result is about 14\% smaller than previously determined with powder samples \cite{Skoulatos2009}.

\subsection{Spin waves}

Inelastic neutron data were collected in  49 constant-$Q$ scans for momentum transfers along several high-symmetry directions in the crystal. The cumulative data are shown in the false color intensity plot in Fig.~\ref{fig:Neutron}. Representative scans are shown in symbols in Fig.~\ref{fig:Neutron_Scan}. Distinct intensity peaks in the latter were attributed to scattering by spin waves. The data were modeled with linear spin wave theory (SWT), as computed using the SpinW library \cite{Toth2015} and assuming a collinear classical CAF ground state as in Fig.~\ref{fig:struct}~b). The key parameters of the model are two nn and two nnn Heisenberg exchange constants. All excitations were assumed resolution limited. The computed scattering intensities were numerically folded with the spectrometer resolution function computed in the Popovici approximation  \cite{Popovici1975} using the software package ResLib  \cite{ResLib}. The magnetic form factor for \VV as well as neutron polarization factor were taken into account. The quasi-elastic background contribution was modeled with an empirical  Lorentzian profile. The additional parameters were an overall scale factor for all scans and a flat background. All the measured energy scans were analyzed in a single global fit yielding a good agreement with $\chi^2=2.50$.  The resulting exchange constants are tabulated in Tab.~\ref{tab:J}. The corresponding dispersion relation is shown in solid lines in Fig.~\ref{fig:Neutron}, with line thickness proportional to the computed intensity. Individual scans, computed using the results of the global fit, are shown in solid lines in Fig.~\ref{fig:Neutron_Scan}. Note that the present measurement is unable to distinguish between the diagonal exchange constants $J_{2,1}$ and $J_{2,2}$, which can be interchanged without any loss of agreement between data and model.

\begin{table}[h]
	%	\begin{center}
	\def\arraystretch{2}
	\begin{tabularx}{\textwidth}{ X  X }
		\hline
		\hline
		$\hspace{0.5cm} J_{1,1} = \SI{-0.35(1)}{\meV}$    & \hspace{0.5cm} $J_{2,1} = \SI{+1.21(1)}{\meV}$  	\\
		$\hspace{0.5cm} J_{1,2} = \SI{-0.42(1)}{\meV}$  	& \hspace{0.5cm} $J_{2,2} = \SI{+0.32(1)}{\meV}$  \\
		\hline
		\hline
	\end{tabularx}
	
	\caption{\label{tab:J} Exchange parameters obtained in a linear spin wave theory fit to the inelastic neutron spectrum measured in \SrZn at $T=0.1$~K. The diagonal exchange constants are only determined up to a permutation  $J_{2,1} \leftrightarrow J_{2,2}$.}
\end{table}

\section{Discussion}

\subsection{A comparison of the three vanadyl phosphates}
The results presented above paint a very curious picture of the magnetism of \SrZn. This material is clearly more similar to \PbV than to \BaCd. Like the former and unlike the latter, it has no alternation of exchange interaction strengths along any directions \footnote{\PbV might have a very small alternation of nn-interactions along one crystallographic direction, but the effect is not manifest in the measured spin wave spectrum.} and correspondingly a simple columnar-antiferromnagnetic ground state. Like in \PbV, nn interactions in \SrZn are very similar along the main crystallographic axes, while the nnn coupling constants along the two diagonals differ by a factor of 3 to 4. The relative strength of AF interactions in \SrZn (${\left< J_2 \right>}/{\left< J_1\right>}=-2.0$) is not drastically smaller than in \PbV (${\left< J_2 \right>}/{\left< J_1\right>}=-2.9$), but much larger than in \BaCd (${\left< J_2 \right>}/{\left< J_1\right>}=-0.6$), where nn-ferromagnetism dominates.

At the same time there are substantial differences: quantum fluctuations in \SrZn are much stronger than in \PbV. It is already evident in the much suppressed value of the ordered moment in the SrZn-system, which is almost as small as in \BaCd \cite{Skoulatos2019}. The same conclusion follows from the correspondence between measured SWT exchange constants in zero field and the value of the saturation field. In the fully polarized phase the single magnon instability field (a lower bound for the actual saturation filed) is directly and exactly expressed through the SWT magnon energy at the ordering wave vector of the low-field structure:
\begin{equation}
\mu_0 H_{\mathrm{sat}} \ge \mu_0 H_{\text{single-magnon}} = \frac{4S}{\mu_0 g} \left( J_{1,1}+J_{2,1}+J_{2,2} \right)
\end{equation}
Substituting the numbers in Table~\ref{tab:J} gives $\mu_0H_\mathrm{sat}\ge20.3$~T, by a third larger than the reported saturation field of \SI{14.2}{\T} \cite{Tsirlin2009}. This implies that the exchange constants  determined in our analysis must be strongly renormalized compared to their actual values. Note that a similar computation for \PbV yields almost exactly the correct saturation field \cite{Tsirlin2009}.

\subsection{Anisotropy}
In both the Pb$_2$- and SrZn- systems the mysterious pre-saturation phase appears only for certain directions of applied field. Magnetic anisotropy is therefore crucial for either stabilizing the new phase in a particular geometry or for suppressing the phase transition in other field orientations. In these $S=1/2$ materials two types of anisotropy need be considered: that of the $g$-tensor and that of the exchange interactions.
\subsubsection{$g$-tensor anisotropy}
Within each V$^{4+}$ layer, the VO$_6$ bipyramids are distinctly buckled. This must give rise to an anisotropic and {\em staggered} gyromagnetic tensor. A staggered $g$ tensor becomes relevant in applied magnetic fields, since it results in a staggered and potentially symmetry-breaking spin field. To analyze this effect in Fig.~\ref{fig:struct} and \ref{fig:struct_Pb}~c) we show the orientations of the local environments of V$^{4+}$ ions within a single layer in the two materials. Each ion is represented by a quiver of three orthogonal arrows. The quivers on equivalent ions are related by the corresponding crystallographic symmetry operations.

For \SrZn, applying a magnetic field along $\mb{c}$ will produce a staggered spin field along the $\mb{a}$ direction. However, the $(0,\pi)$-modulation and direction of this field coincides with those of the CAF structure. A similar conclusion can be drawn for \PbV. If, however, a magnetic field is applied along either $\mb{a}$ or $\mb{b}$, it generates a staggered spin field along $c$. That staggered field is directed transverse to the zero-field CAF spin alignment. More importantly, it has a different periodicity compared to the CAF state, namely $(\pi,\pi)$ or Neel-type. We note that any additional pre-saturation phase transition is absent whenever this Neel field along $\mb{c}$ is present. An intriguing possibility is that it is precisely this field that destroys the pre-saturation state that would otherwise be manifested. This would also explain the similarity in the behavior observed in the two vanadates. A detailed theoretical study will be required to clarify this issue.

\subsubsection{Exchange anisotropy}
In the Moriya-Anderson superexchange mechanism, for a $S=1/2$ system, any anisotropy of exchange interactions is due to antisymmetric  Dzaloshinskii-Moriya interactions and to the so-called KSEA term, an easy axis parallel to the Dzyaloshinskii vector  \cite{Kaplan1983,Shekhtman1992,Shekhtman1993}. Unfortunately, the pattern of Dzyaloshinskii vectors in layered vanadyl phosphates is generally very complex. In all three  materials mentioned in this work none of the in-plane bonds feature inversion centers or symmetry planes. As a result, symmetry allows for an arbitrarily oriented Dzyaloshinskii vector $\mb{D}$ accompanying every Heisenberg exchange constant $J$. The consequences of this complex microscopic picture are difficult to elucidate. The best we can do is try to at least guess the strength of this anisotropy from bulk measurements. For a classical XXZ antiferromagnet $(J_{zz}-J_{xx})/J_{xx}\sim (H_\mathrm{SF}/H_\mathrm{sat})^2$. Naively applying this to \SrZn gives $D/J\sim H_\mathrm{SF}/H_\mathrm{sat}\sim 0.05$, quite typical for 3d transition metal system.

\section{Conclusion}
We have shown that \SrZn shows a pre-saturation phase similar to that in \PbV. Qualitative and quantitative differences between the two materials suggest that this behavior may be generic. Either its origin or its suppression in certain field geometries may be related to a complex pattern of magnetic anisotropy in these compounds.

\section{Acknowledgments}
This work is partially supported by the Swiss National Science Foundation under Division II. This work was additionally supported by the Swiss State Secretariat for Education, Research and Innovation (SERI) through a CRG-grant. 

\bibliography{SrZn-bib}

\end{document}